\documentclass[10pt,a4paper,twoside]{article}

\usepackage{indentfirst}			
\usepackage{graphicx}				
\usepackage{amsgen,amsfonts,amssymb,amsbsy}	

\newenvironment{resum}{\begin{quote}\small}{\end{quote}}

\newcommand{\bfsf}[1]{\textsf{\textbf{#1}}}

\begin{document}

\thispagestyle{plain}		

\begin{center}


{\LARGE\bfsf{Purely Magnetic Spacetimes}}

\bigskip


\textbf{Alan Barnes}


\textsl{University of Aston, Birmingham, B4 7ET, U.K.}

\end{center}

\medskip


\begin{resum}
Spacetimes in which the electric part of the Weyl tensor vanishes (relative to some timelike unit vector field) are said to be purely magnetic. Examples of purely magnetic spacetimes are known and are relatively easy to construct, if no restrictions are placed on the energy-momentum tensor.  However it has long been conjectured that purely magnetic vacuum spacetimes (with or without a cosmological constant) do not exist. The history of this conjecture is reviewed and some advances made in the last year are described briefly.   A generalisation of this conjecture first suggested for type D vacuum spacetimes by Ferrando and S\'aez is stated and proved in a number of special cases.  Finally an approach to a general proof of the conjecture  is described using the Newman-Penrose formalism based on a canonical null tetrad of the Weyl tensor.
\end{resum}
\smallskip

\vspace{-5 pt}
\section{Introduction}
\noindent The electric and magnetic parts of the Weyl tensor with respect to some unit timelike vector field $u^a$ are defined by
\begin{equation}
E_{ab} = C_{acbd}u^c u^d \qquad H_{ab} = C^*_{acbd}u^cu^d
\label{eq.1}
\end {equation}
respectively, where $C^*_{abcd}$ is the dual of the Weyl tensor. Using this decomposition, which was first introduced (for the vacuum Riemann tensor) by Matte \cite{matte}, the Bianchi identities take a form analagous to Maxwell's electromagnetic field equations.  Spacetimes for which $H_{ab}=0$ are said to be {\it purely electric\/} whilst those in which $E_{ab}=0$ are said to be {\it purely magnetic\/}.
If the electric and magnetic parts of the Weyl tensor  are proportional, that is if 
\begin{equation}
\nu E_{ab} = \mu H_{ab}
\label{eq.2}
\end{equation}
for some scalar fields $\nu$ and $\mu$ (not both zero), we will say that the electric and magnetic parts are {\it aligned\/}. The aligned case, of course, includes the purely electric and magnetic fields as the special cases $\nu=0$ and $\mu=0$ respectively.  In all aligned cases the complex tensor $Q_{ab}=E_{ab} + i H_{ab}$ is a complex multiple of a real symmetric tensor and so may be diagonalised by a tetrad rotation leaving $u^a$ fixed.  Thus the Petrov type is I, D or O (see, for example \cite{exact-sol}) and the eigenvalues $\alpha_A$ of $E_{ab}$ and $\beta_A$ of $H_{ab}$ are also proportional, that is $\nu \alpha_A = \mu \beta_A$ where uppercase Latin indices run from 1 to 3.  The purely electric and magnetic cases correspond to $\beta_A = 0$ and $\alpha_A = 0$ respectively.  Although the condition in (\ref{eq.2}) appears to depend on choice of a vector field $u^a$, in fact this is not the case.  If $u^a$ is such that (2) holds then it is a Weyl principal vector. Thus for Petrov type I, $u^a$ is, like the rest of the Weyl principal vectors, determined uniquely (up to sign) by $C_{abcd}$ and in the type D case it is determined  up to a boost in the plane of the repeated principal null directions of the Weyl tensor.

There is an alternative characterisation of type I fields satisfying (\ref{eq.2}): namely that the four principal null directions of the Weyl tensor are linearly dependent and span the three-dimensional vector space orthogonal to the eigenvector of $Q_{ab}$ corresponding to the eigenvalue 
$\lambda_A = \alpha_A + i\beta_A$ of smallest absolute value \cite{trump, nar, mcint}.  For type I fields the Weyl tensor invariant
$M = I^3/J^2 - 6$ is real and positive or infinite, where 
\begin{equation}
I = \lambda_1^2 + \lambda_2^2+\lambda_3^2 \qquad 
J  = \lambda_1^3 + \lambda_2^3+\lambda_3^3 = 3\lambda_1\lambda_2\lambda_3
\label{eq.3}
\end{equation}
Thus the Petrov type is I($M^+$) or I($M^\infty$) in the extended Petrov classification of Arianrhod \& McIntosh \cite{arian}.  In fact in the vacuum case (with or without a cosmological constant) a result first proved by Szekeres \cite{szek}, but usually attributed to Brans \cite{brans}, shows that the case  I($M^\infty$), that is where one of the eigenvalues of $Q_{ab}$ is zero, cannot occur.

For non-vacuum spacetimes some authors use the term {\it purely electric\/} to mean 
$R^*_{acbd}u^c u^d = 0$.  As shown in \cite{trump}, this condition is equivalent to the two conditions: $H_{ab} =0$ and $u^a$ is a Ricci eigenvector.  For vacuum spacetimes ($R_{ab} = \Lambda g_{ab}$), the two definitions of purely magnetic are equivalent.  Almost invariably in studies of the non-vacuum case, $u^a$ is assumed to be a Ricci eigenvector and so there is little danger of confusion arising from the two different definitions of the term  {\it purely electric\/}.  This is not so for the purely magnetic case which some authors define as $R_{acbd}u^c u^d = 0$.  This condition does not imply $E_{ab} = 0$ unless the Ricci tensor satisfies
\begin{equation}
R_{ab} = u_{(a}q_{b)} - u^c q_c g_{ab}
\label{eq3.5}
\end{equation}
for some vector field $q_a$.  In fact, any two of the conditions $R_{acbd}u^b u^d = 0$, $E_{ab} = 0$ and (\ref{eq3.5}) imply the third \cite{hadd}.  The non-equivalence of these two definitions of  {\it purely magnetic\/} has led to some confusion in the literature; see \cite{arian2} for a discussion of this.  Note that even for vacuum spacetimes, the two definitions are not equivalent unless the cosmological constant $\Lambda$ vanishes.  In this paper  the term {\it purely magnetic\/} will always be used to mean $E_{ab} = 0$.

Many examples of purely electric spacetimes, both vacuum and non-vacuum, are known in the literature. For example all static spacetimes are necessarily purely electric \cite{EK} as are all shear-free and hypersurface othogonal perfect fluid spacetimes \cite{barn1}. However, relatively few purely magnetic spacetimes are known although the situation has improved in recent years (see for example \cite{loz, carmin}).  To date none of the purely magnetic solutions that have been found, satisfy the vacuum field equations: namely $R_{ab}= \Lambda g_{ab}$.  This has led researchers to conjecture that there are no purely magnetic vacuum spacetimes (excluding the trivial constant curvature case).  A number of special cases of this conjecture have been proved.  For example Hall \cite{hall} showed that there were no purely magnetic type D vacuum metrics and this result was rediscovered in \cite{mcint}.  Note that Hall also proved a related result: namely that there are no vacuum type II spacetime in which the eigenvalues $\lambda_A = \alpha_A +i \beta_A$ of $Q_{ab}$ are purely imaginary.  For type I spacetimes the conjecture has been proved under the additional assumption that the vector field $u^a$ is shear-free by Barnes \cite{barn2} and this result was rediscovered by Haddow \cite{hadd}.  More recently van der Berg \cite{vdB1, vdB2} has proved the conjecture in a further two special cases: namely when the vector field $u^a$ is either hypersurface orthogonal ($\omega^a = 0$) or geodesic ($\dot u^a = 0$).
\vspace{-5 pt}

\section{A Generalised Conjecture}

\noindent Recently Ferrando \& S\'aez \cite{FS1} considered special type D fields in which 
\begin{equation}
E_{ab} = R H_{ab}
\label{eq.4}
\end{equation} 
where $R$ is a real constant.  This class of spacetimes is clearly a specialisation of the aligned case defined in (\ref{eq.2}).  Alternatively this class may be characterised by the assumption that the eigenvalues $\lambda_A$ of $Q_{ab}$ are all real multiples of a single complex constant or equivalently that all these eigenvalues have the same constant argument.   For conciseness below these fields will be referred to as having {\it constant argument\/}.  Note that the purely magnetic fields are a special case of the constant argument fields with $R=0$, but that the purely electric case is excluded (it corresponds informally to $R = \infty$).  We will also use the term {\it constant argument} for Petrov type II fields in which the eigenvalues of $Q_{ab}$ have constant argument although (\ref{eq.4}) is not now valid.

In \cite{FS1} it was shown that there are no vacuum type D fields of constant argument.  This leads one to consider whether there are any Petrov type I or II vacuum fields of constant argument.  Below it is shown that 
\newcounter{N}
\begin{list}
{\Roman{N}.}{\usecounter{N}
       \setlength{\rightmargin}{\leftmargin}}
\item there are no vacuum constant argument  Petrov type I fields in which the vector field $u^a$ is shear-free;
\item  there are no vacuum constant argument Petrov type II fields.  
\end{list}
\noindent These two results generalise those for the purely magnetic case in \cite{barn2, hadd} and  \cite{hall, mcint} respectively.   Furthermore,  in the interval between the conference and the appearance of these proceedings, Ferrando \& S\'aez \cite{FS2} have shown that there are no vacuum constant argument Petrov type I fields in which the vector field $u^a$ is hypersurface orthogonal; this is a direct generalisation of the result for the purely magnetic case in \cite{vdB1}. This naturally leads to the conjecture that there are no vacuum fields whatsoever  of constant argument. 

The proof of result I above is now presented.  Putting $\sigma_{ab}=0$ in the vacuum  Bianchi identities (4.21a) and (4.21c) of  \cite{ellis}, one obtains
\begin{equation}
h^a_b E^{bc}_{\ \ ;d}h^d_c + 3H^a_b  \omega^b   =   0 \qquad
h^a_b H^{bc}_{\ \ ;d}h^d_c - 3E^a_b  \omega^b   =   0
\end{equation}
where $h_{ab} =g_{ab}+u_a u_b$ is the projection tensor into the three-space orthogonal to the vector field $u^a$ and $\omega^a$ is the vorticity vector of $u^a$. Using (\ref{eq.4}) it follows that 
\begin{equation}
(1+ R^2) H^a_b \omega^b = 0
\end{equation} 
and thus either $\omega^b$ is zero or it is an eigenvector of $H_{ab}$ (and hence of $Q_{ab}$) with zero eigenvalue. The latter case is excluded by the theorem of Szekeres \cite{szek, brans} discussed in the Introduction. In the former case the congruence defined by $u^a$ is shear-free and hypersurface orthogonal and so by a well-known result $H_{ab} = 0$ (see for example \cite{barn1, exact-sol}), hence $E_{ab} =0$ and the spacetime is conformally flat and so not of type I.  In both cases a contradiction is obtained and the required result I is proved.

\section{An Approach using the NP Formalism}

\noindent In this section vacuum constant argument spacetimes will be investigated using the Newman-Penrose (NP)  formalism \cite{NP, exact-sol}.  When the timelike vector field $u^a$ satisfies certain kinematic restrictions, an orthonormal tetrad approach is natural and has been used in many previous investigations of purely magnetic and constant argument spacetimes (for example \cite{barn1, hadd, vdB1, vdB2, FS1, FS2}).  However, when there are no {\it a priori\/} restictions on the vector field $u^a$, then a null tetrad formalism also becomes attractive.

Suppose $(u^b, e^b_A)$ is an orthonormal Weyl principal tetrad of the spacetime. If we introduce an associated null tetrad $(k^a, l^a, m^a, \bar m^a)$ defined by 
\begin{equation}
k^a = 1/\sqrt{2}(u^a+e_3^a) \qquad l^a = 1/\sqrt{2}(u^a-e_3^a) \qquad m^a = 1/\sqrt{2}(e_1^a+i e_2^a)
\end{equation}
then, if the spacetime is of Petrov type I or D, the NP Weyl curvature components satisfy (see \cite{exact-sol} p.\ 51)
\begin{equation}
\Psi_1=\Psi_3 = 0 \qquad \Psi_0 =\Psi_4 = (\lambda_2 -\lambda_1)/2 \qquad \Psi_2 = -\lambda_3/2
\end{equation}
where  the $\lambda_A$'s are the eigenvalues of $Q_{ab} = E_{ab}+iH_{ab}$.  For Petrov type D, without loss of generality, 
$\lambda_1 =\lambda_2$ and so $\Psi_0=\Psi_4 = 0$.  
For Petrov type II we have instead
\begin{equation}
\Psi_0 = \Psi_1=\Psi_3 = 0 \qquad \Psi_4 = -2 \qquad \Psi_2 = -\lambda_3/2
\end{equation}

From (9) \& (10) for Petrov type I, II \& D fields, a null tetrad may be chosen so that $\Psi_1=\Psi_3 = 0$.  In such a frame the vacuum Bianchi identities reduce to (see \cite{exact-sol} p.\ 81) 
\begin{eqnarray}
\bar \delta \Psi_0 & = & (4\alpha - \pi)\Psi_0+3\kappa\Psi_2 \\
\Delta \Psi_0 & = & (4\gamma-\mu)\Psi_0 +3\sigma \Psi_2 \\
D \Psi_4 & = & (\rho - 4\epsilon) \Psi_4 - 3\lambda \Psi_2 \\
 \delta \Psi_4 & = & (\tau - 4 \beta) \Psi_4 -3 \nu \Psi_2 \\
D \Psi_2 & = & - \lambda \Psi_0 + 3\rho \Psi_2 \\
\Delta \Psi_2 & = & \sigma \Psi_4 -3\mu \Psi_2 \\
\bar \delta \Psi_2 & = & \kappa \Psi_4 - 3\pi \Psi_2 \\
\delta \Psi_2 &= & -\nu \Psi_0 +3 \tau \Psi_2
\end{eqnarray}
\vspace{-16 pt}  

We may now proceed to prove proposition II of the previous section: namely that there are no vacuum constant argument spacetimes of Petrov type II (or of type D).  For constant argument type II and D fields, the NP Weyl tensor components satisfy $\Psi_0 = \Psi_1=\Psi_3 = 0$ and $\Psi_2 = A\exp(iB)$ where $A > 0$ is a real scalar field and $B$ is a real constant (and $B \neq 0, \pi$ as the purely electric case is excluded). With these assumptions it is easy to deduce from the Bianchi identies (11--18) that
\begin{eqnarray}
&\kappa=\sigma=\tau+\bar \pi=0 \qquad & \rho = \bar \rho \qquad \mu =\bar \mu \\
&D A  = 3\rho A \qquad & \Delta A= -3\mu A
\end{eqnarray}
\vspace{-16 pt}  

\noindent With the restrictions (19) on the spin coefficients the commutator relation (7.6a) of \cite{exact-sol} becomes
\begin{equation}
(\Delta D - D \Delta) = (\gamma + \bar \gamma) D + (\epsilon + \bar \epsilon) \Delta
\end{equation}
Applying this commutator to $A$ it may be deduced with the aid of (20) that 
\begin{equation}
\Delta \rho + D \mu = (\gamma + \bar \gamma)\rho -(\epsilon + \bar \epsilon)\mu
\end{equation}
but from the Ricci identities (7.21h) \& (7.21q) of \cite{exact-sol},  on making use of 
(19), it follows that
\begin{equation}
\Delta \rho + D \mu = (\gamma + \bar \gamma)\rho -(\epsilon + \bar \epsilon)\mu +A(e^{iB}-e^{-iB}) 
\end{equation}
It follows immediately that either $A=0$ or that $B= 0, \pi$. In either case we have a contradiction and so vacuum type D and type II spacetimes of constant argument cannot exist.  This proof extends the result of \cite{hall} from the purely magnetic to the constant argument case and extends the result of \cite{FS1} for the constant argument case from type D to type II fields. The proof is somewhat more direct than the original proofs of these two results and is valid without modification when the cosmological constant $\Lambda$ is non-zero.

For type I vacuum fields of constant argument the NP Weyl tensor components satisfy $\Psi_1=\Psi_3 = 0$, $\Psi_0= \Psi_4 =A_0\exp(iB)$ and $\Psi_2 = A_2\exp(iB)$ where $A_0$ and $A_2$ are positive scalar fields and $B \neq 0, \pi$ is a real constant.  The Bianchi identities (11--18) then lead to a number of algebraic relations, quadratic in the spin-coefficients, which must be satisfied. Currently work is in progress investigating the integrability conditions of these algebraic relations in the hope of proving the conjecture that  no vacuum constant argument and/or purely magnetic fields can exist or of integrating the NP equations to find some counter-examples to these conjectures.
\vspace{-10 pt}

\end{document}